\pdfoutput=1
\documentclass[a4paper, 11pt]{JINST}

\usepackage{textcomp}
\usepackage{amsmath}
\usepackage{lineno}

\title{GPS-based CERN-LNGS time link for Borexino}

\author{
B.~Caccianiga$^{1}$, 
P.~Cavalcante$^{2}$,
G. Cerretto$^3$,
H. Esteban$^4$,
G.~Korga$^{5,6}$, 
M.~Misiaszek$^{7}$,
M.~Orsini$^{2}$, 
M.~Pallavicini$^{8}$, 
V. Pettiti$^3$,
C. Plantard$^3$,
A.~Razeto$^{2}$\footnote{alessandro.razeto@lngs.infn.it}

{\scriptsize $^1$Dipartimento di Fisica, Universit\`a di Milano and INFN Milano, via Celoria 16, I-20133 Milano, Italy}

{\scriptsize $^2$Laboratori Nazionali del Gran Sasso, SS 17bis Km 18+910, I-67010 Assergi (AQ), Italy}

{\scriptsize $^3$Optics Division, INRIM (Istituto Nazionale di Ricerca Metrologica), Torino, Italy.}

{\scriptsize $^4$Time Department, Real Instituto y Observatorio de la Armada (ROA), San Fernando, Spain.}

{\scriptsize $^5$Physics Department, University of Houston, Houston, TX 77004, USA}

{\scriptsize $^6$MTA Wigner FK RMI, (Wigner Research Centre for Physics), 1121 Budapest, Hungary}

{\scriptsize $^7$Institute of Physics, Jagiellonian University, ul. Reymonta 4, PL-30059 Krakow, Poland}

{\scriptsize $^8$Dipartimento di Fisica, Universit\`a di Genova and INFN Genova, via Dodecaneso 33, I-16146 Genova, Italy}

}

\abstract{
We describe the design, the equipment, and the calibration of a new GPS based 
time link between CERN and the Borexino experiment at the Gran Sasso Laboratory in Italy. 
This system has been installed and operated in Borexino since March 2012, and used for a precise measurement
of CNGS muon neutrinos speed in May 2012. The result of the measurement will be reported in a different letter. }


\begin{document}

\section{Introduction}
\label{sec:intro}

This paper describes new equipment recently installed in the Borexino experiment for an accurate measurement of the CNGS \cite{bib:cngs} muon neutrinos ($\nu_\mu$) speed. Particularly, we report about the design, installation, and performance of a new system, the High Precision Timing Facility (HPTF), a GPS-based timing facility with a calibrated time-link to the CERN GPS receiver and with continual real time monitoring of the time delay to the underground laboratory.

This system is ready and has also been made available to other LNGS experiments\footnote{The LVD and Icarus experiments are using it.}. It has been already used to measure the muon neutrino speed in May 2012, during a special short bunch run of the CNGS beam. The system performed as expected. The result of the speed measurement will be reported in a different letter.

The Borexino detector is extensively described elsewhere \cite{bx08det,bxmuondet,bxfilling,bib:old1,bib:old2}. We recall here, however, the most relevant elements with the purpose of keeping this paper as much as possible self-consistent. 

\begin{figure}[t]
\begin{center}
\includegraphics[width=0.8\textwidth]{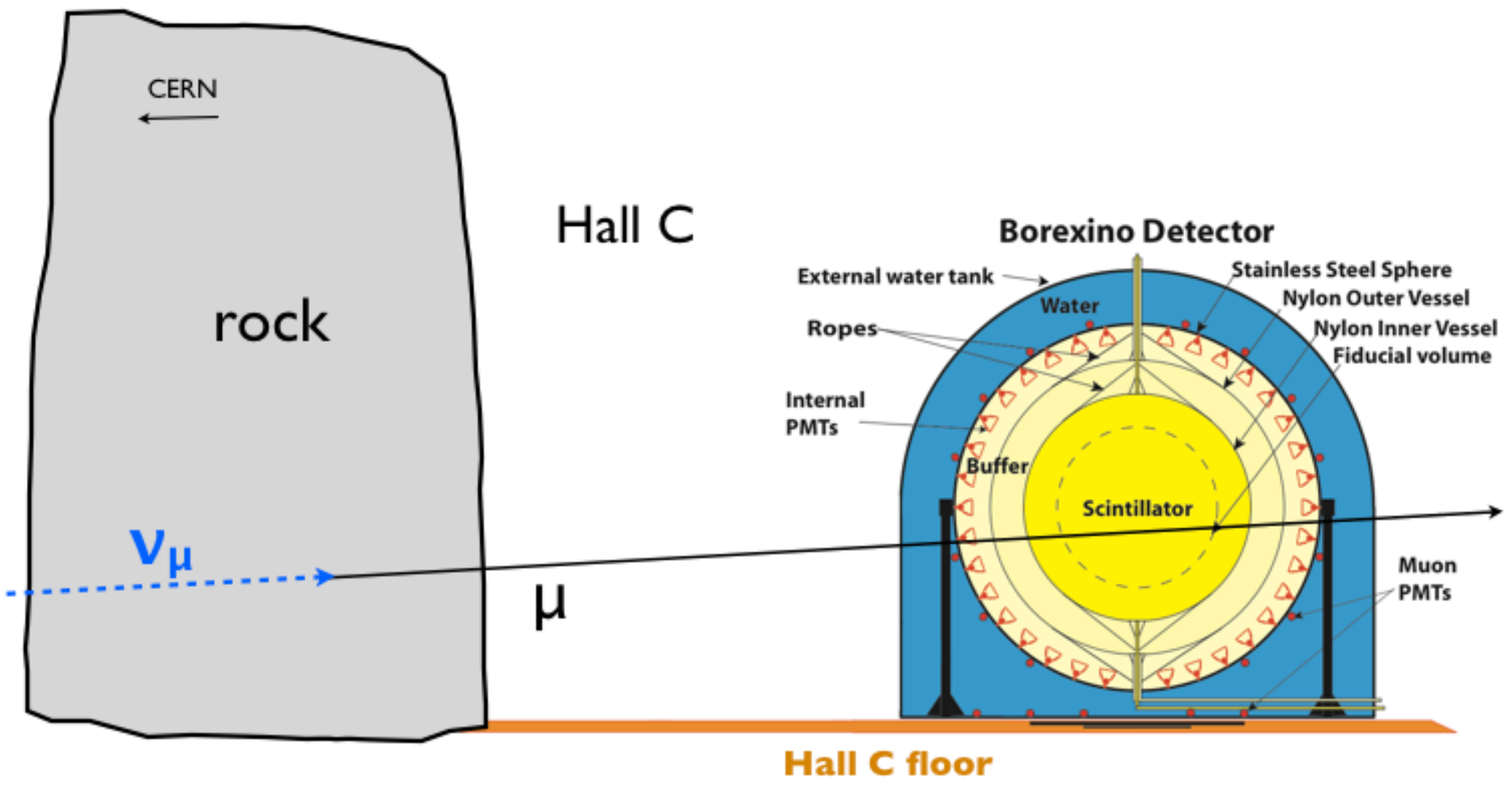}
\end{center}
\caption{Schematic drawing of the Borexino detector inside an even more schematic Hall C at Gran Sasso. Most of the CNGS events are muons produced by neutrinos in the rock upstream. CERN is on the left. Muons are inclined 3.2$^\circ$ above the horizontal axis. }
\label{fig:detector}
\end{figure}

As far as CNGS events are concerned, Borexino is made of two independent parts (see Fig. \ref{fig:detector}): a large domed steel tank of 18 m diameter and 16.9 m height filled with 2100 t of ultra-pure water and instrumented with 208 PMTs which detect the muon Cherenkov emission; a Stainless Steel Sphere (SSS, external radius 6860 mm) filled with $\approx$1300 m$^3$ of liquid scintillator and buffer liquid viewed by 2212 PMTs, which detect the scintillation light emitted both by the true scintillator (yield $\approx$ 500 p.e./MeV) and by the buffer liquid (yield $\approx$ 25 p.e./MeV). The active medium for the detection of CNGS muons in the SSS is therefore the whole Stainless Steel Sphere volume. The muons crossing the Water Tank only are not used in this analysis, but the information provided by the muon detector is used to reconstruct the entrance point of the muons in the SSS.

The CNGS muon neutrinos are detected in Borexino through their interaction with an atomic nucleus, which may be in the detector (internal event) or, much more frequently, in the rock upstream (external event, see Fig. \ref{fig:detector}). Neutral current interactions in the rock do not provide a visible signal in Borexino, so most of the events are quasi-horizontal muons that cross both the Water Tank (WT) and the Stainless Steel Sphere (SSS). The much less abundant internal events are detected both via charged and neutral current interactions.

Muons produce a clear and fast signal in both detectors. In both cases, the light is collected by 8'' ETL 9531 PMTs which were selected also because of their low time jitter (1.3 ns). Through high quality underwater cables (equalized within 250 ps), the signal reaches the Front End electronics. At this point the two detectors are made in a different way and we refer to \cite{bx08det} and \cite{bxmuondet} for more details.

The standard Borexino trigger \cite{bib:triggerlaben} is made with custom digital modules that are able to compute the number of (k) photomultipliers which fired within a time window of 99 ns. The calculation is made by an FPGA every 33 ns. When the number k exceeds a programmable threshold, a Digital Signal Processor initiates the triggering sequence by latching and reading an ESAT-100 GPS slave and by sending appropriate signals to Borexino electronics. 

This architecture has proved to be perfectly adequate to solar neutrino physics because it allows operating the detector with a very low energy threshold (25 fired PMTs in typical runs, corresponding to about 50 keV in deposited energy in the scintillator). However, the jitter introduced by the FPGA + DSP chain limits the precision with which the absolute time of an event can be measured by this system. We have measured this jitter by firing the timing laser with a sufficiently large pulse to trigger the detector, and we have obtained 32 $\pm$ 0.1 ns, in nice agreement with expectations. Besides, the ESAT-100 GPS slave rounds the event Universal Time Coordinated (UTC) time information to 100 ns, so its intrinsic accuracy is of the order of 100 ns / $\sqrt{12}\simeq$ 29 ns.

We have therefore installed an additional system which computes the total charge of one event by adding linearly the analogue signal of each photomultiplier. This system is made with passive resistor networks and linear amplifiers, and has a total gain of 1.12 mV/p.e.; a standard NIM linear discriminator (Lecroy 623BLZ, Leading Edge Discriminator) with a threshold of 90 mV provides the final analogue trigger. This new system has a much lower jitter. The higher energy threshold that can be safely set to this new system is not a problem for CNGS muons, which typical deposit into the detector much more than 10 MeV. Here and in the following we call CNGS\_Trigger\_ID the trigger generated by this system.

The Outer Detector (OD) trigger is described in \cite{bxmuondet}. The noise level in the detector, mostly due to a residual amount of light that we could not completely remove from the system because of the very large number of cable and pipe feedthroughs, does not allow to operate the OD with an analogue trigger. We therefore decided to use the standard OD trigger even for the CNGS run. The jitter of this trigger is higher than the one provided by the new inner detector system (about 30 ns), but it is not used in this analysis for timing information, so it is not critical.

A custom module performs the logical OR of the inner detector and outer detector triggers (K1 module in Fig. 2), providing what we call a CNGS\_Trigger signal. The module has an internal well calibrated fan-out which provides two copies of this signal within a maximum jitter of 100 ps. The module has also an internal monostable which prevents the generation of a double pulse when both CNGS\_Trigger\_ID and CNGS\_Trigger\_OD signals are present (which is usually the case for muons crossing the SSS, either from CNGS or not). A delay of 1 $\mu$s is tuned between the two signals, so that it is possible to distinguish the two also by simply using the time of an event. 

The purpose of the HPTF described in this work is to measure with high precision the time--difference between a CNGS\_Trigger signal and the moment in which the proton beam has hit the target at CERN. 

\section{The High Precision Timing Facility}
This section is devoted to the main scope of this paper, which is the description of the design guidelines of the High Precision Timing Facility (HPTF).

The system can deliver the Borexino triggers from the Hall C underground up to the core of the facility which is located outside, close to the GPS antenna, and measure with high precision the time of a given trigger. It is composed of several parts: a connection from underground to external laboratory with real time monitoring of the propagation delay, and GPS receiver coupled to a low jitter Rb clock, and a set of Time Interval Counters and high precision fan out modules for timing measurements and signal distribution. The next sections describe each of these parts. Section \ref{sec:results} show the performance obtained with this system.

\subsection{Optical connection to HPTF}
\label{sec:optics}
One of the main feature of this system is the ability to bring the triggering signal generated by the detector close to the GPS system via optical fibers and, even more important, to monitor on a real time basis the length of the link.

\begin{figure}[t]
\begin{center}
\includegraphics[width=\textwidth]{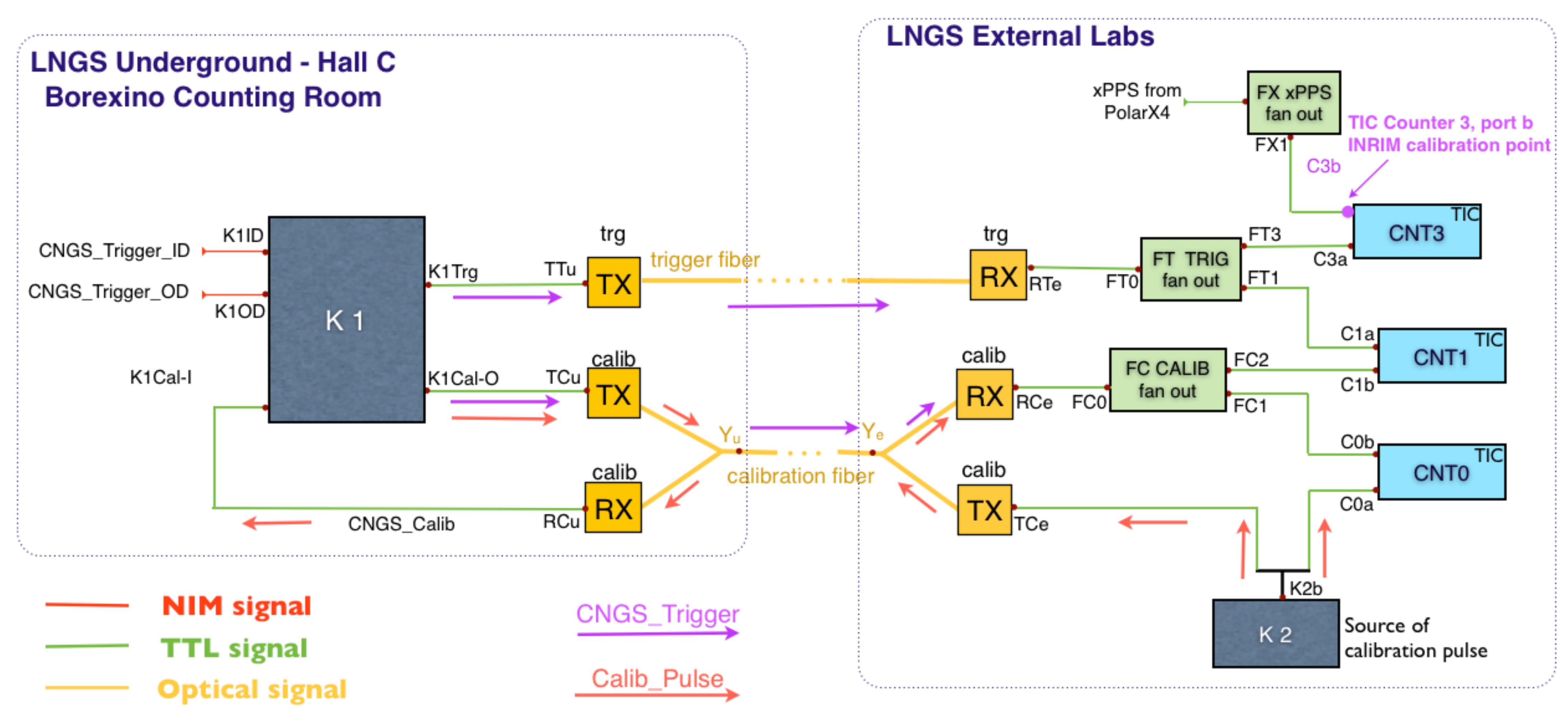}
\end{center}
\caption{Layout of the connection system between the Borexino counting room (left part) and external laboratory (right part). The optical fiber connections (yellow lines) are approximately 9 km long. The system delivers the trigger signal to the External Laboratory with very small jitter ($<$100 ps) and monitors in real time the fiber length within an accuracy better than 100 ps. The purple lines show the paths of the trigger along the two fibers; the red lines show the path of the calibration pulse back and forth along the calibration fiber. See text for detailed description. }
\label{fig:optical}
\end{figure}
This is done by means of the optical fiber link shown in Fig. \ref{fig:optical}, which is made of two fibers: a main fiber (used to deliver the trigger), and a calibration fiber (used as well to deliver a second copy of the trigger and to bring back and forth a calibration pulse). 
The custom module shown on the left in Fig. \ref{fig:optical} (K1) provides two copies of the CNGS\_Trigger signal: one of them is converted into an optical signal and sent out to the external laboratory through one optical fiber (main fiber); a second copy is put in logical OR with the calibration signal that comes from a second fiber (calibration fiber) and sent back to the same fiber by means of a Y-shape optical connection and a pair of RX-optical receiver and TX-optical transmitter.  

The conversion from TTL signal to optical is made with a low jitter J724-13 RX optical converter from Highland Technology\footnote{http://www.highlandtechnology.com/DSS/J724DS.shtml}. A similar device, J730-13, converts from optical to TTL (TX). The typical jitter of these units is below 12~ps (for the RX-TX couple), more than sufficient for our purpose. 
The transceivers need to be compatible with single mode fibers and allow a power loss of about 12~dB, to cope with the LNGS fibers from external labs to underground (10~km fiber, 6~dB attenuation) plus the fiber splitter for the calibration fiber (2-3~dB attenuation depending on direction).

In the external laboratory the optical signal are converted back to TTL pulses by the same transceivers. As shown in Fig. \ref{fig:optical}, the CNGS\_Trigger signal coming from the main fiber is fan-out in two copies: one is used to measure the time difference between the trigger itself and the Pulse Per Second (PPS) signal coming from the GPS receiver; a second copy is used to measure the time difference with the other copy of CNGS\_Trigger that has traveled through the calibration fiber.

A calibration pulse is generated every 10 s by a second custom module (K2), sent through the calibration fiber to the underground laboratory and then brought back through the same fiber. The time difference between the starting calibration pulse and the come back pulse is measured.

\begin{figure}[t]
\begin{center}
\includegraphics[width=0.9\textwidth]{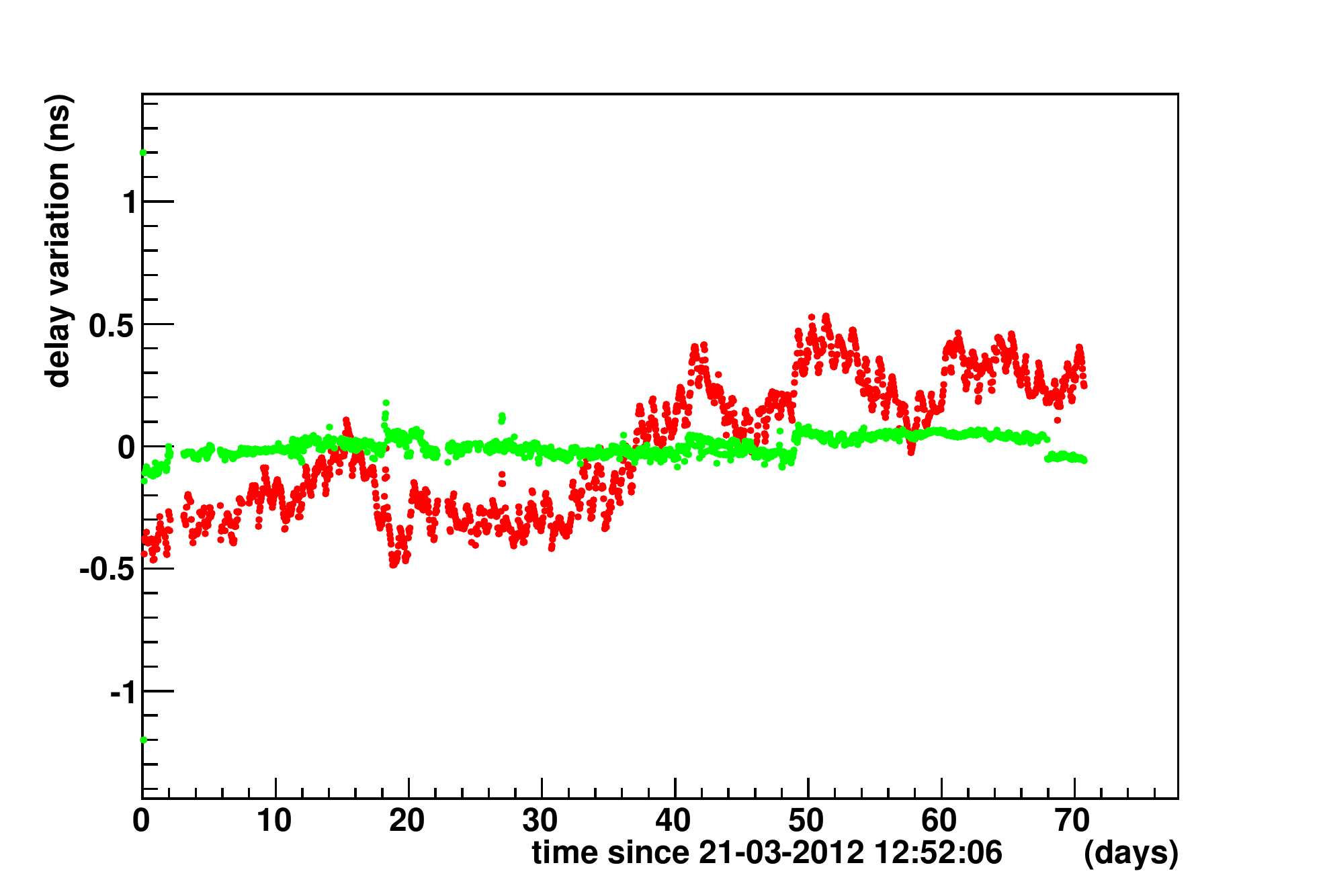}
\end{center}
\caption{The relative time variation of the optical fiber lengths over a period of about 70 days: the red curve is the variation with respect to its own mean value of the length of the loop (sum of two fibers); the green is the variation with respect to its own mean value of the difference between the same two fibers. 
The loop shows day-night small variations (a few 100 ps) and long term variation due to temperature changes (about 1 ns in two months). The difference is much more stable and never changes more than a few hundreds ps. The real time monitoring of both quantities allows to measure the instantaneous delay between external labs and hall C within 100 ps or less. }
\label{fig:fiberdelays}
\end{figure}

This configuration of the system allows to measure both the time difference of the two optical paths (using the normal CNGS\_Trigger pulses that are synchronously sent along both ways) and the total delay of the sum of the calibration and main fibers. Using the fact that all the delays related to the electrical connections both in the Hall C counting room and in the HPTF room in the external laboratory have been very carefully measured, this redundant double fiber system allows to measure accurately and to monitor the stability of the optical fibers delays. As an example, we show in Fig. \ref{fig:fiberdelays} the result of about 70 days of data for time difference of the main path and of the calibration path and for the calibration path length. Day-night effect is clearly visible in the fiber length at the level of a few 100 ps and the total variation in two months is about 1 ns. The difference of the two paths is stable within a few 100 ps along the whole period. 

The precision with which the two paths are measured every 10 min is 100 ps or better. 
All time difference measurements are done using Time Interval Counters that are described later in section \ref{sec:tics}.

\subsection{Clock and signal distribution}
\label{sec:clocks}
A high stability, low jitter clock with 10~MHz frequency is required as both source for the GPS receiver and as a time base for the Time Interval Counters (see later). A minimum intrinsic accuracy of $10^{-9}\ s/s$ is required, to allow 1~ns precision over 1~second measure. A phase noise better than 100~dBc is required as well to keep the time jitter below 1~ns.

\begin{figure}[h]
\begin{center}
\includegraphics[width=0.8\textwidth]{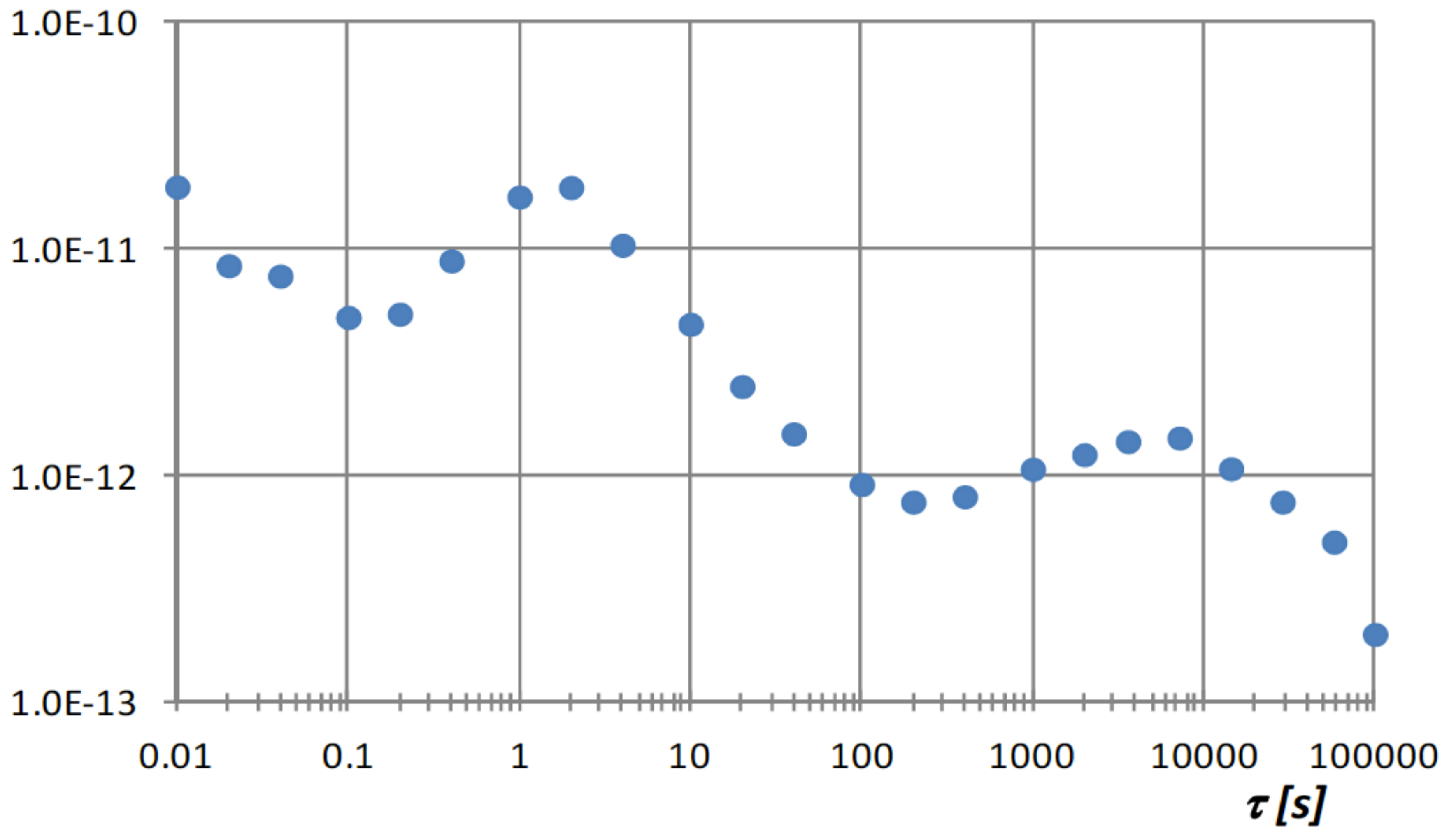}
\end{center}
\caption{Allan deviation of the GPS--locked Rb clock measured over a period of one week. The Allan deviation is the square root of Allan variance, a measure of the frequency stability in clocks, oscillators and amplifiers. }
\label{fig:allanRb}
\end{figure}

The Rubidium reference source FS725 from Stanford Research\footnote{http://www.thinksrs.com/products/FS725.htm} (Rb) provides such features. Moreover, the oscillator can be locked with an external GPS receiver. This additional feature is very useful, because in such configuration both the oscillator frequency
and the phase are continually corrected with a time constant of 8~hours, providing a long term stability that is much better than the one achieved with the Rb clock only. 
The instability of the Rb clock in this configuration was measured at INRIM\footnote{Istituto Nazionale di Ricerca Metrologica, italian institute of metrology} and is about 1.0~10$^{-11}$ s/s ($\tau$=1s).
The FS725  provides 18 sinusoidal 10~MHz outputs plus 4 PPS signals that are obtained by division from the main outputs; in case the device is locked with an external signal, the oscillator is kept in phase with it.
The auxiliary Timing GPS receiver used to lock the Rb is 
M12M-T\footnote{www.rabel.org/archives/Motorola\_Oncore/m12m\%20users\%20guide\%201.0.pdf} 
which provides a low jitter Pulse Per Second (PPS), about 10~ns, more than sufficient for our purpose. Fig. \ref{fig:allanRb} shows the Allan deviation of the 
Rb clock frequency as a function of time measured at INRIM over a period of one week (the observation time must be at least 3-4 times
the measuring time).  

In HPTF several logic pulse distributors are needed. In order to minimize the error in the measurements and the jitters, we have used only high quality, low jitter, low temperature drift, and thermally compensated channels to distribute the critical signals, and particularly the PPS signal from PolarX4, the main Borexino trigger and the calibration pulses.

The PD5-RM-B pulse distribution amplifier from SpectraDynamics\footnote{www.spectradynamics.com/manuals/PD5-RM-B\%20Operating\%20Inst\%20110\%20VAC.pdf}  provide such characteristics: 
the units are 1~U 19" Rack form factor providing each 2 independent distribution amplifiers with 5 output each;
our qualification measurements, done at the INRIM facilities in Torino, show that the first 4 channels of each amplifier are aligned in time within 100~ps, while the fifth is within 200~ps; the temperature coefficient is about 3~ps/$^\circ$C.

\subsection{Time Interval Counters}
\label{sec:tics}
Time Interval Counters (TICs) are commercial devices that measure the time difference of two digital signals ('start' and 'stop') with high accuracy. They perform therefore the same function of the more familiar TDCs, but with the capability to measure long time intervals (up to 1 s) with high precision (up to 12 decimal digits). 

The main uncertainty to the measurement of long time intervals is due to the stability of the internal clock. For this reason, for most of the models the clock can be provided externally. In our case, one copy of the 10~MHz Rb clock described in section \ref{sec:clocks} is delivered to each TIC. 

In the HPTF we have used 5 TICs produced by Pendulum\footnote{http://www.spectracomcorp.com/ProductsServices/TestandMeasurement/FrequencyAnalyzersCounters/ \\ $~~~~~~~~$/CNT90TimerCounterAnalyzer/tabid/1280/Default.aspx}, 3 of model CNT-90 (accuracy 100~ps) and another 2 of model CNT-91 (accuracy 50~ps); among other features, these devices can be programmed to provide a relative timestamp for each measurement,
which is the time of the start since the last reset, with the instrument resolution. The readout is done via USBTMC a protocol for instrumentation over USB.

\subsection{Time Link and GPS receivers}
\label{sec:link}

The time offset between the Cesium clock at CERN and the GPS disciplined Rubidium clock installed in the HPTF at Gran Sasso is an important contribution to the error of the neutrinos speed measurement and must therefore be known accurately. 
In order to evaluate such a time difference, the GPS system has been used. 

Time and Frequency Transfer techniques based on geodetic GPS receivers measurements are, indeed, among the most useful tools for the remote comparison of atomic clocks and time scales, and are currently used worldwide in combination with the TWSTFT (Two-Way Satellite Time and Frequency Transfer) system, for the realization of UTC and TAI (International Atomic Time) time scales by the BIPM (Bureau International des Poids et Measures)~\cite{bib:utc3}.  

The basic idea of this approach is to use the pseudo-range measurements (i.e. the raw distance between the ground receiver and orbiting satellites antennas) produced by dedicated geodetic GPS receivers physically connected to the atomic clocks (through the clock 1PPS and Frequency electrical signals) to be synchronized with respect to an intermediate common time scale, usually the GPS Time or a time scale having a known reference to it. 

In order to achieve this result, the pseudo-range measurements generated by the GPS receivers have to be processed with dedicated algorithms, implementing proper geophysical models and solutions that allows to get the required timing measurement with acceptable characteristics in terms of precision and stability. 
Among these algorithms, the P3 and the PPP (Precise Point Positioning) can be used~\cite{bib:utc4}\cite{bib:utc5}. The first one compensates the ionosphere effect, while the second one, in order to further improve the precision, processes receivers measurements along with IGS (International GNSS Service) precise satellite orbits/clocks and applies special models for site displacement (e.g. Solid Earth Tides, Ocean Loading) and satellite attitude effects (e.g. Satellite Antenna offsets, Phase Wind-Up Corrections)~\cite{bib:utc6}. 

\begin{figure}[t]
\begin{center}
\includegraphics[width=0.95\textwidth]{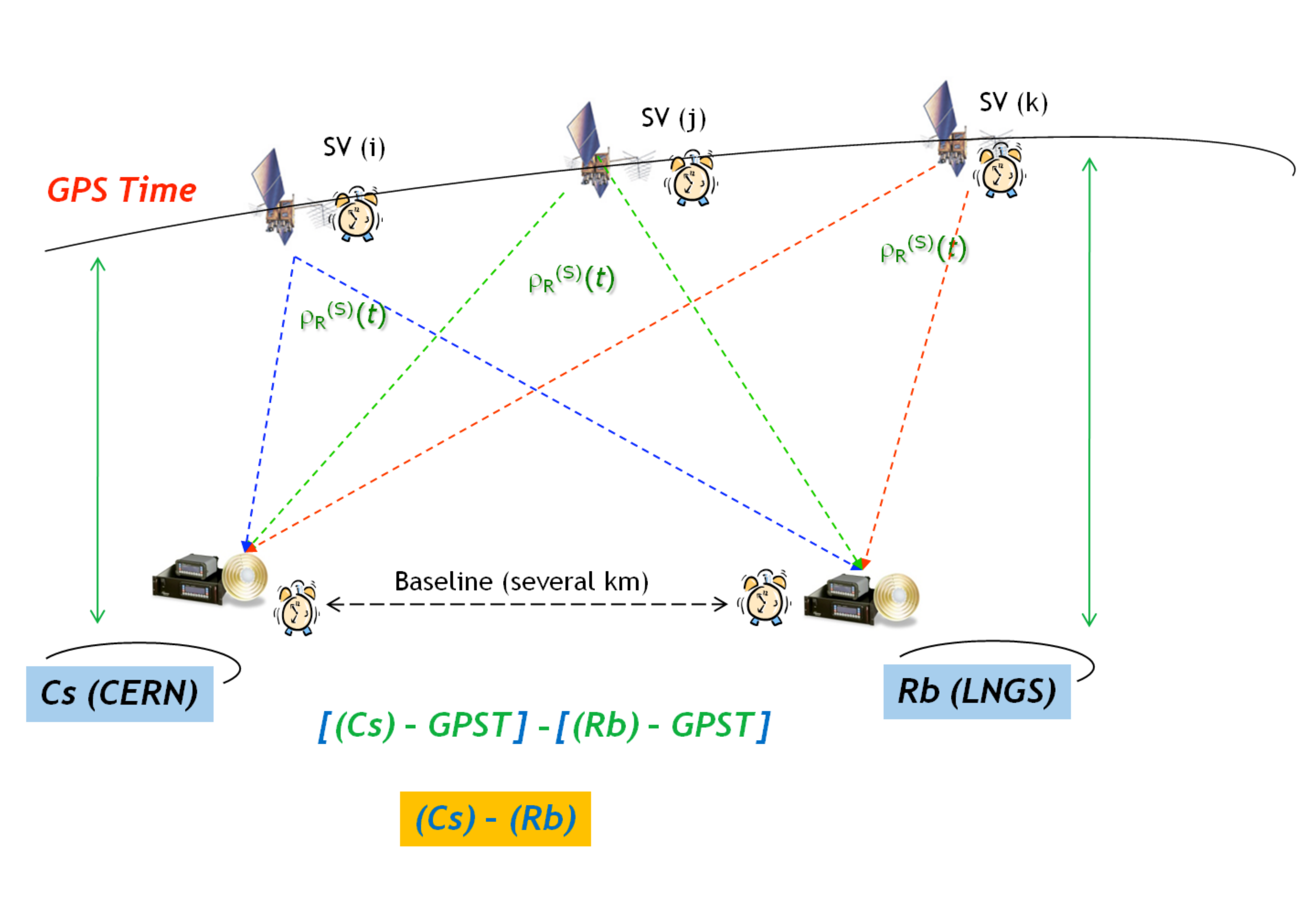}
\end{center}
\caption{Clocks comparison approach with GPS.}
\label{fig:gps-layout}
\end{figure}

In Fig. \ref{fig:gps-layout}, a scheme showing the clocks comparison approach using GPS is reported, having indicated with SV(i) each of the satellites in view and with $\rho$ the pseudo-distance between them and the receiver.

Despite the good performances in term of precision achieved, in order to provide accurate time transfer by means of a GPS link (like the CERN-LNGS one), it is necessary to carry out calibrations, to be repeated periodically to verify the long term stability of the equipments.  

The calibration of the GPS time link between CERN and LNGS has been carried out in "link mode"~\cite{bib:utc7}. In particular, this approach involves the receivers at both laboratories (a PolaRx2eTR at CERN and a PolaRx4TR at LNGS, both produced by Septentrio), and a reference or traveling receiver (TR, namely a DICOM GTR50) that is circulated between the two Laboratories and is set in "common clock" and "near zero baseline" (antennas very close) set-up with the local receivers. 

This GPS receiver is not continuously available at LNGS, but it has been provided by INRIM in order to perform the required calibration. 
The GPS link calibration value (GPSCAL) for the couple of receivers hosted at CERN and LNGS Laboratories is calculated by the simple difference of the so called common clock difference results (in the following, $\Delta_{CERN}$ and $\Delta_{LNGS}$), using the ionosphere free P3 data, generated with the R2CGGTTS V5.0 software developed at the time section of the Royal Observatory of Belgium \cite{bib:utc4}, and the PPP data, generated with the NRCan PPP 1087 software developed at the Geodetic Survey Division of Natural Resources Canada \cite{bib:utc5}. 

In simplified mathematical terms, this can be expressed as:

$\Delta_{CERN} = T_{GPS}(TR) - T_{GPS}(R_{CERN}) $   $~~~~~~$ @ CERN

$\Delta_{LNGS}  = T_{GPS}(TR) - T_{GPS}(R_{LNGS})$  $~~~~~~~$ @ LNGS

$\delta_{12} = \Delta_{LNGS} - \Delta_{CERN}$ 

where $R_{LNGS}$ is the GPS receiver at Gran Sasso, $R_{CERN}$ is the receiver at CERN, TR is a traveling receiver and the $T_{GPS}$ values are the time difference measurement between the receivers clocks and the GPS Time, intending them averaged over a certain period, typically one day. Hence, the difference between CERN and LNGS clocks can be stated as in the following expression:

$$K_{CERN-LNGS} = T_{GPS}(R_{CERN}) - T_{GPS}(R_{LNGS}) - \delta_{12}$$ 		

\begin{figure}[t]
\begin{center}
\includegraphics[width=0.95\textwidth]{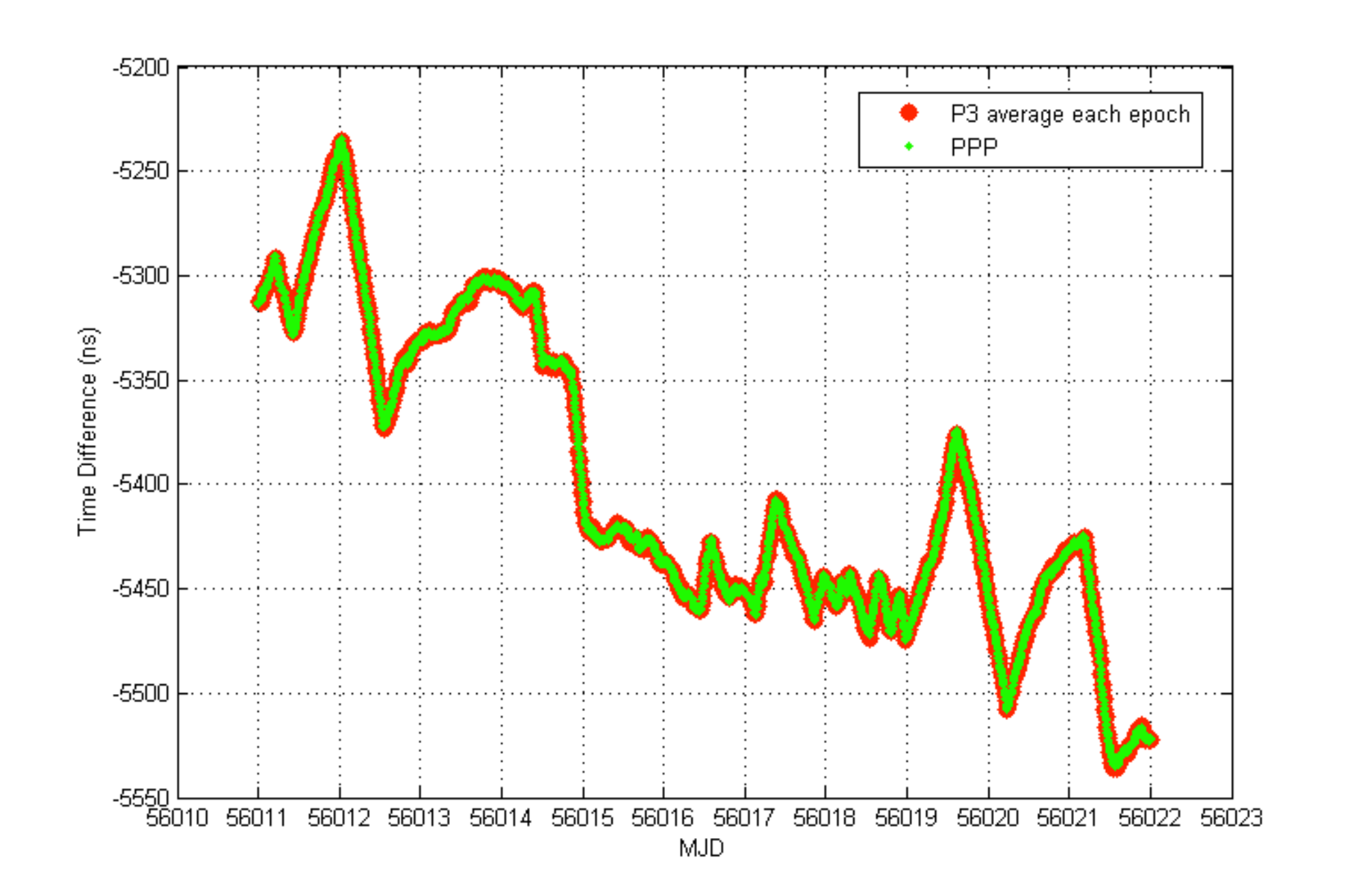}
\end{center}
\caption{CERN and LNGS clocks time differences as estimated with P3 and PPP algorithms for the period MJD 56011-56021, corresponding to 2012 March 25$^{th}$ - April 4$^{th}$.}
\label{fig:p3-ppp}
\end{figure}

\begin{figure}[t]
\begin{center}
\includegraphics[width=0.85\textwidth]{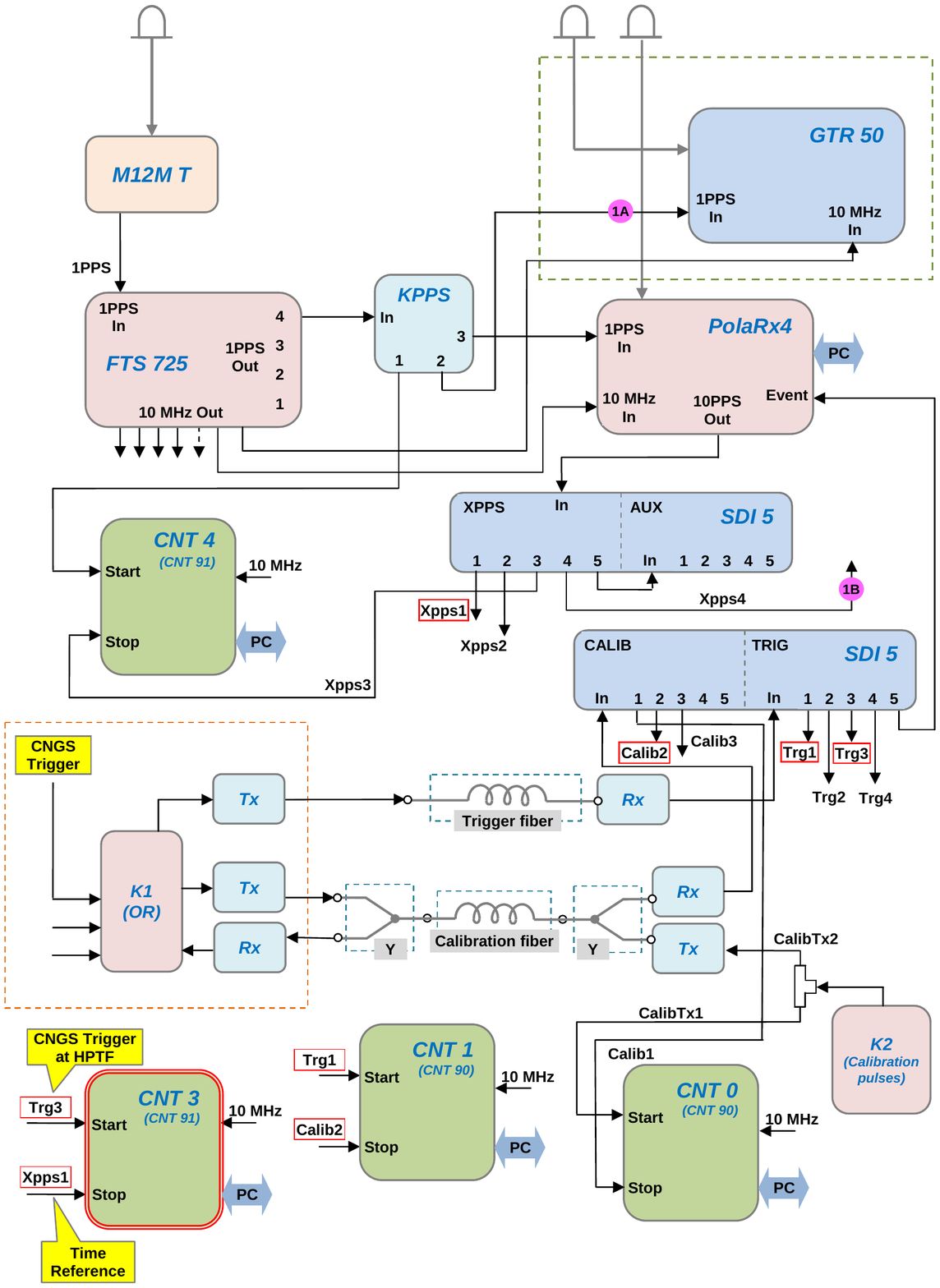}
\end{center}
\caption{Schematic layout of the High Precision Timing Facility at LNGS. In the figure, the meaning of the symbols is the following: 
M12M T: single frequency/ multichannel GPS receiver; GTR50: double frequency/multichannel GPS receiver; FTS 725:	GPS disciplined rubidium oscillator; KPPS: pulse distribution amplifier; PolaRx4: double frequency/multichannel GPS receiver; 
SDI 5: double pulse distribution amplifier. }
\label{fig:LayoutINRIM}
\end{figure}

The calibration $K_{CERN-LNGS}$ of the CERN-LNGS GPS link was estimated in (232.6$\pm$1.1)~ns for P3, and basically the same for PPP. Resulting uncertainties are slightly higher than 1 ns, as expected for this type of calibration [15]. 

It is assumed that this calibration value remains valid until any change or event happens in any of the two installations and it can thus be included in the calculation of the time differences between CERN and LNGS reference clocks. 
Once the GPS link calibration has been computed, an automatic processing system has been implemented to daily compute the GPS link corrections, intended as the calibrated (corrected) time difference between the reference clocks at CERN and LNGS. 

This system performs the common view of P3 files obtained processing the RINEX "observation" and "navigation" files [16] generated at both sites. For each day, a time series up to 90 points has been selected, nominally performed every 16 minutes (epochs). Previously, a MAD (Median of the Absolute Deviation) based filtering procedure is performed at each epoch, and finally a 3$\sigma$ dynamic filter with respect to the daily Vondrak smoothing is applied, with the aim of removing any outlier. 

The time difference between CERN and LNGS reference clocks has been also computed applying the PPP algorithm, yielding 5 minutes estimates, using IGS "rapid" precise satellite orbits and clocks. For each day a similar 3$\sigma$ dynamic filter is applied. PPP algorithm has been configured to yield also 30 seconds estimates, using IGS "final" ephemeris and 30 seconds clocks products. 

With this further solution, the uncertainty derived from interpolation for any epoch can been reduced significantly. In 
Fig.~\ref{fig:p3-ppp}, an example of calibrated time difference between CERN and LNGS clocks is shown, as computed by both the P3 and the PPP algorithms.

The neutrino speed measurement was done just a few weeks after the calibration, so we expect that is fully reliable. It will be re-done as cross check in the near future. 

The complete layout of all instrumentation and the location of the calibration point in HPTF are shown in Fig. \ref{fig:LayoutINRIM}.

\begin{figure}[t]
\begin{center}
\includegraphics[width=0.85\textwidth]{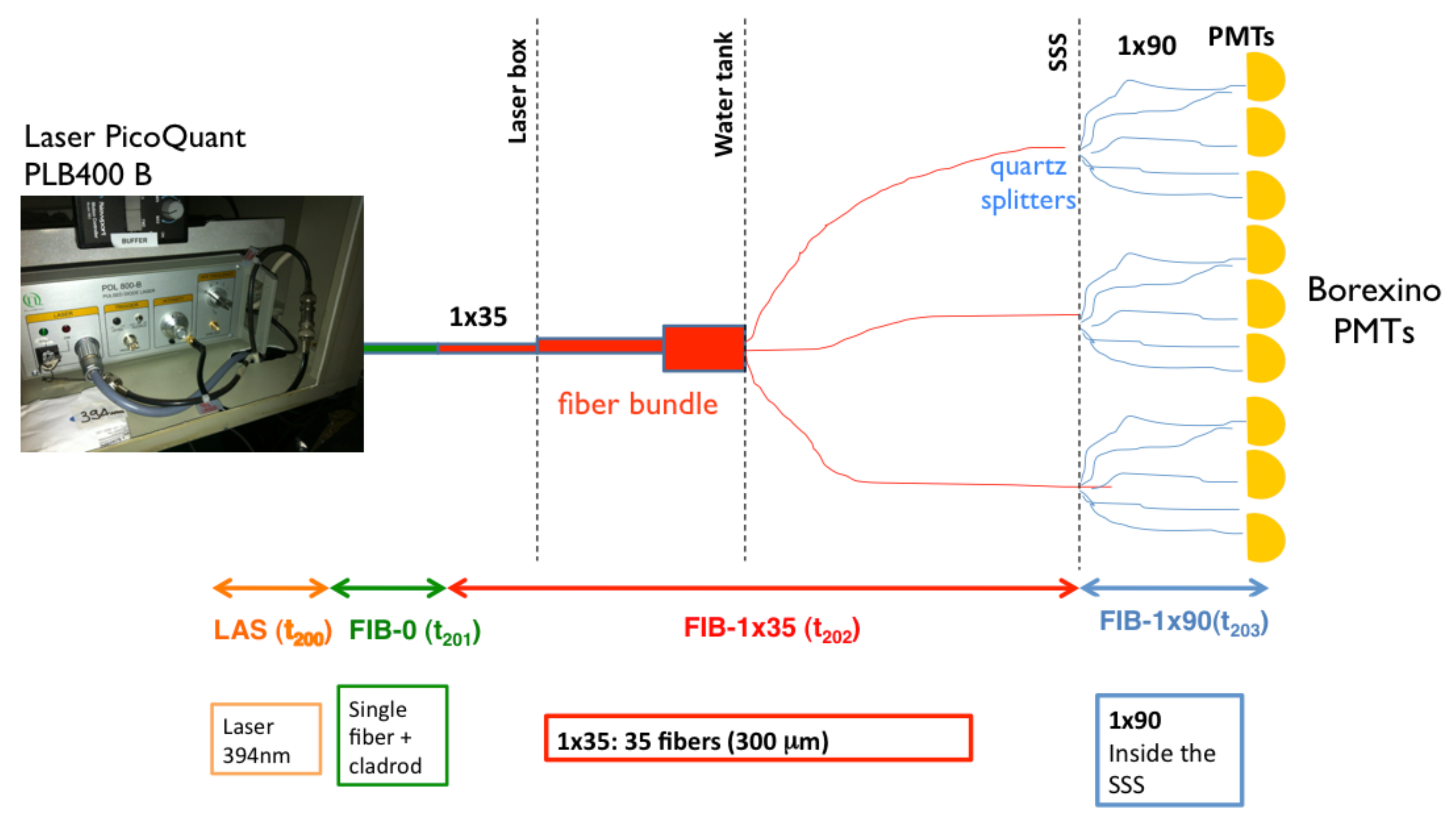}
\end{center}
\caption{Schematic layout of the laser system used to synchronously pulse all inner detector PMTs. }
\label{fig:laser}
\end{figure}

\section{Calibrations and system performance}
\label{sec:results}
We have carefully measured all internal delays relevant for the determination of the neutrino speed. 
The main tool for this is a fast triggerable laser made by Picoquant (model PLB400-B) whose pulse can be split and sent synchronously to all Borexino internal detector PMTs (see Fig. \ref{fig:laser}). 

\begin{figure}[t]
\begin{center}
\includegraphics[width=0.8\textwidth]{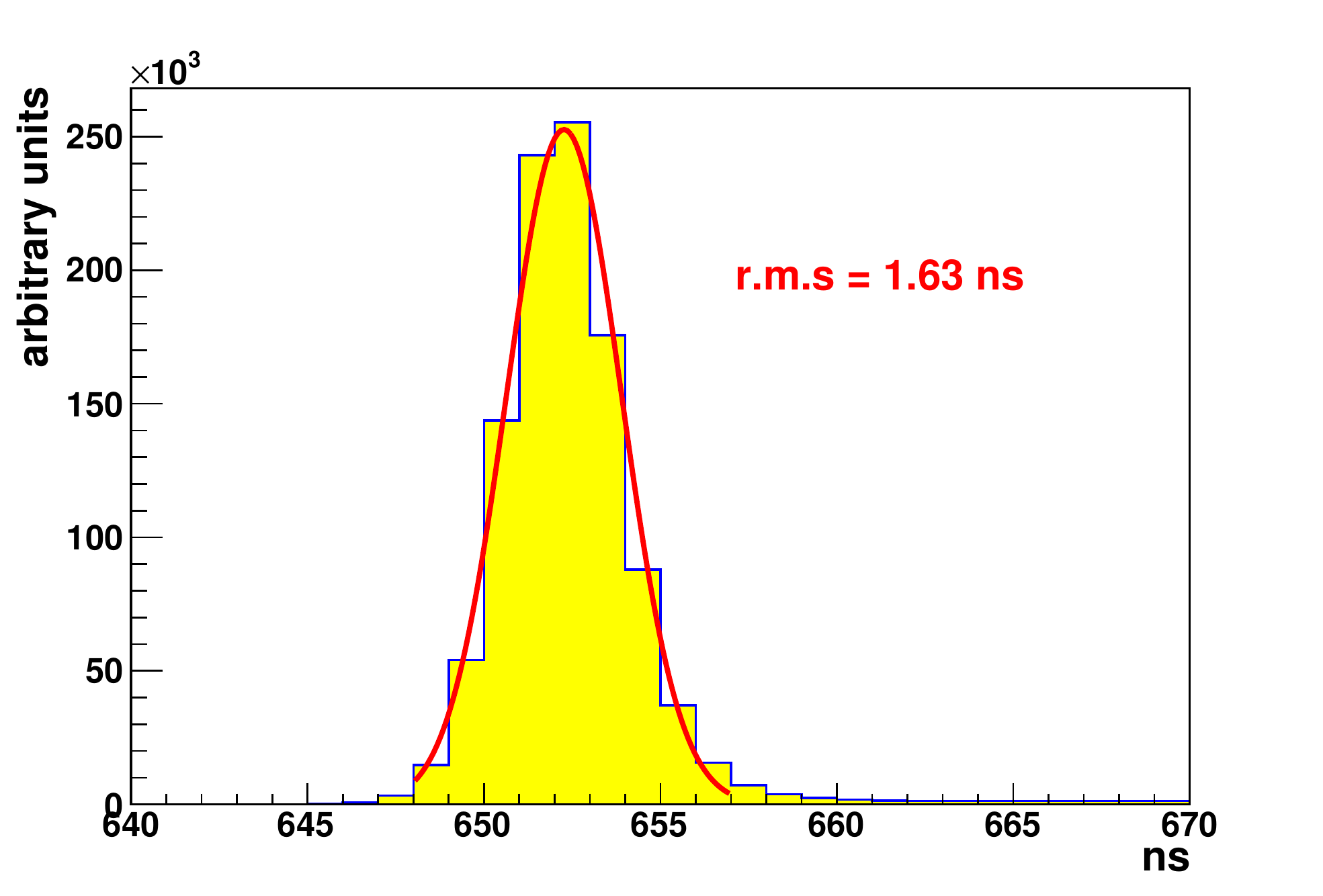}
\end{center}
\caption{Distribution of the photon detection time of all Borexino photomultipliers obtained in a laser run. The distribution
is in good agreement with the expected value due to intrinsic PMT jitter and electronics resolution. The broad non-gaussian tail on the
right is due to light reflection within the SSS and is not included in the fit. The plot is just an example of the capability of the system
to guarantee very good timing calibration.  }
\label{fig:laser2}
\end{figure}

The laser can generate short pulses (70 ps long) at the nominal wavelength of 394 nm, a value at which the Borexino scintillator is very transparent. 
The laser pulse is sent to a bundle of 35 fibers each of them is then split into 90 individual fibers. Each of these fibers reach one PMT inside the sphere. Fig. \ref{fig:laser} shows a schematic layout of the connection between the laser and the Borexino PMTs. 
Fig. \ref{fig:laser2} shows an example of the time distribution of laser hits on all photomultipliers with respect to the laser trigger. The distribution is 1.63 ns wide, in good agreement with the expected width due to intrinsic PMT jitter and electronics resolution.

This system was designed to perform a careful time equalization of the PMTs (a key feature needed
to reconstruct the event position accurately) but it can also be used to generate triggers and measure the propagation delay of the signal through the detector. In order to do so, we had to carefully measure the total delay introduced by the fibers. 

In principle, the length of the fibers can be measured in a standard way by means of an OTDR, Optical Time Domain Reflectometer. However, these instruments usually work at visible or near infrared wavelength, while our timing laser has a wavelength of 394 nm. 

In order to get the desired value, we adopted the following procedure: we have prepared a sample fiber, made exactly of the same material and coating of our fibers (a piece of this fiber was available in our lab from the original Borexino installation).  This sample fiber has been measured geometrically, using the OTDR at 850 nm, and using Borexino system electronics. The measurement was done by placing the reference fiber in the Borexino timing system, and then measuring the delayed pulse by means of the Borexino data acquisition itself. The measurement has an error of 0.5 ns, including systematic effects due to the laser itself (obtained from manufacturer's specifications). 

We have also measured with sub-ns precision all the delays internal to the HPTF system and to the custom module K1 in Fig. \ref{fig:optical}. In this way, from the measurement of the fiber loop and fiber difference in 
Fig.  \ref{fig:optical} we can extract on real time basis the length of the fibers and the delay introduced
by the 10 km long connection from the input to K1 up to the TIC that measures the phase with respect to the Rubidium PPS. 

\begin{figure}[t]
\begin{center}
\includegraphics[width=0.7\textwidth]{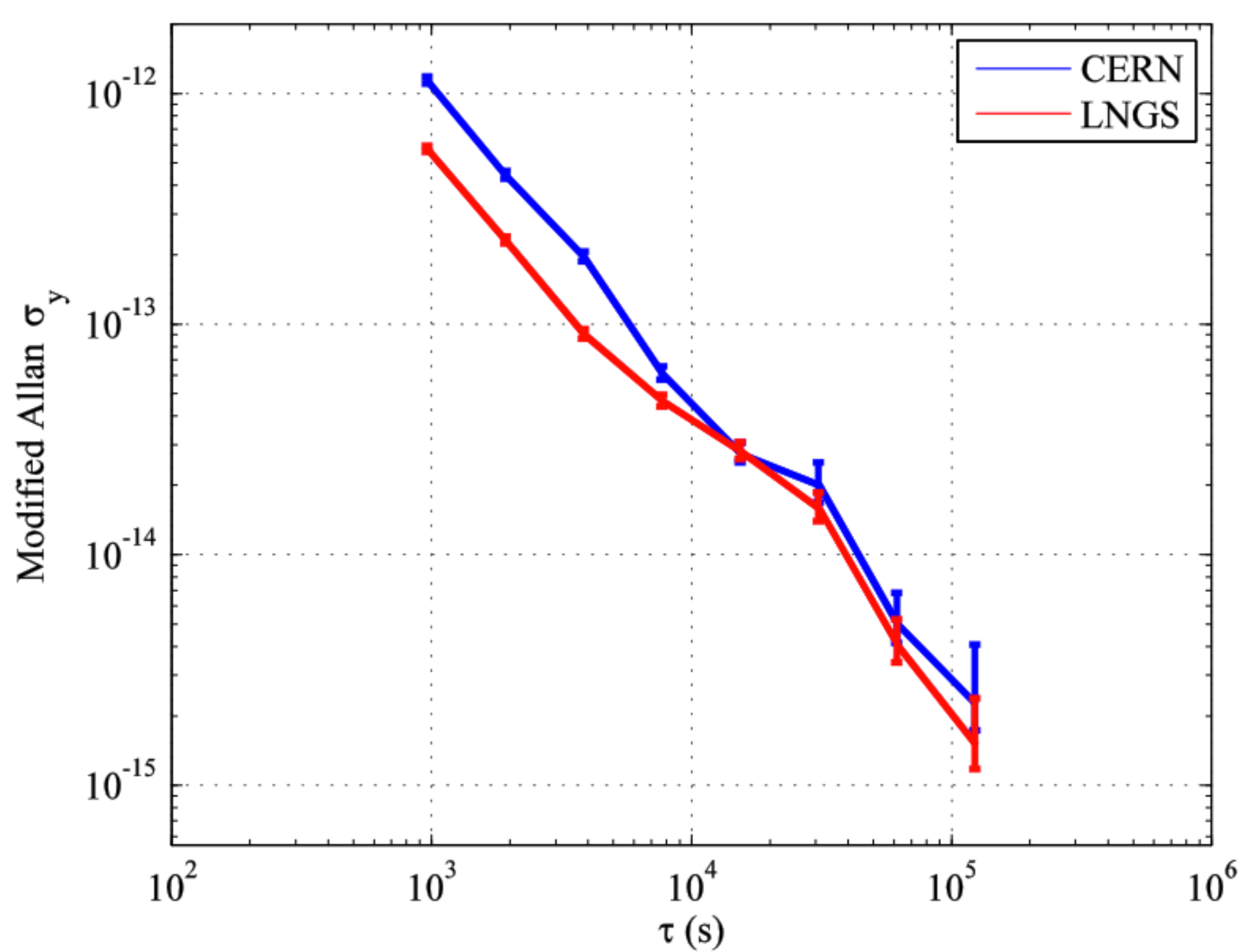}
\end{center}
\caption{Modified Allan deviation for the Common Clock Deviations at CERN and at Gran Sasso.}
\label{fig:allan-cern-lngs}
\end{figure}

As mentioned in the previous section, the GPS time link between CERN and LNGS has been calibrated with an uncertainty 
to be at level of 1.1 ns, in agreement with what achieved for the calibration of the links between receivers hosted at Time and Frequency Laboratories in the frame of the computation of the TAI and UTC time scales by BIPM. 

\begin{figure}[th]
\begin{center}
\includegraphics[width=0.70\textwidth]{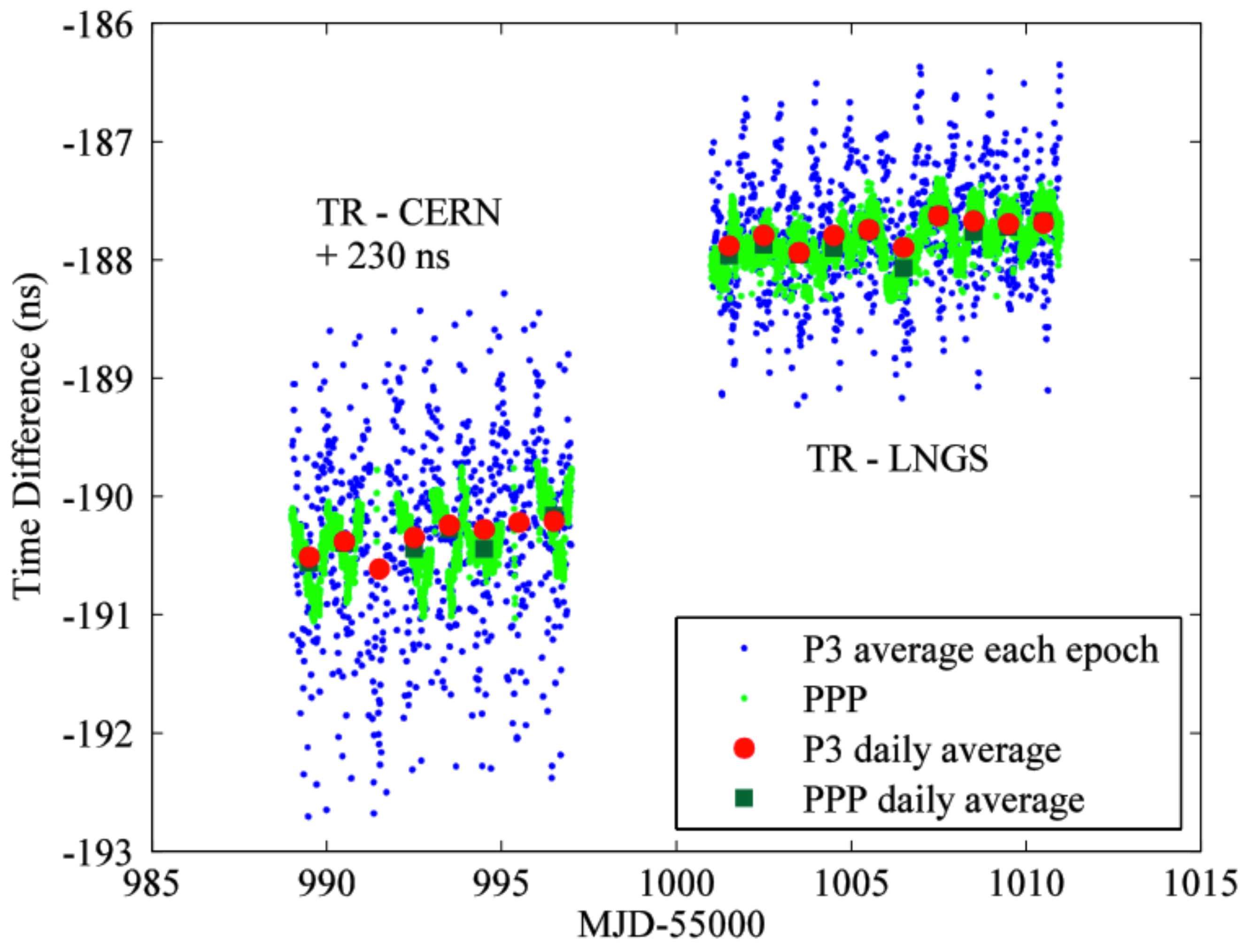}
\end{center}
\caption{P3 and PPP Common Clock Deviations for GPS receivers at CERN ($\Delta_{CERN}$ in the text) and at Gran Sasso ($\Delta_{LNGS}$). The data taking periods are shown in
MJD-55000 and correspond to the periods March 3$^{rd}$-10$^{th}$, 2012 at CERN and March 15$^{th}$-24$^{th}$, 2012 at Gran Sasso.}
\label{fig:calib}
\end{figure}

The calibration has been performed using a Traveling Receiver, normally running at INRIM and driven by the Italian 
Realization of UTC - namely UTC(IT) - operated at CERN between MJD 55989 to 55996 (2012 March 3$^{rd}$-10$^{th}$), 
at LNGS between MJD 56001 to 56010 (2012 March 15$^{th}$-24$^{th}$) and, finally, shipped back to INRIM to 
carry out the "closure" measurements with the initial set-up. Fig. \ref{fig:allan-cern-lngs} shows the results, in terms of CCD as 
computed by P3 and PPP algorithms (estimates every 960 s and 30 s, respectively) and their daily average, while in Fig. \ref{fig:calib} the stability of such differences is reported 

As already shown above, we have also proved the feasibility of an automatic calibration procedure that takes into account all the delay contributions, and guarantees repeatability of the measurements below 0.5 ns.

\section*{Acknowledgements}
We gratefully acknowledge the generous support of the Laboratori Nazionali del Gran Sasso. We also recognize the technical and financial support provided by the LVD and Icarus collaborations, and particularly by W. Fulgione and G. Bruno. 
We also thank and acknowledge the Geodetic Survey Division (GSD) of Natural Resources Canada (NRCan), for providing the PPP software and the "Time Section" of "Royal Observatory of Belgium" (ROB) for providing the P3 software. Special thanks to F. Lahaye (NRCan) and P. Defraigne (ROB) for the kind support and helpful advices provided on the usage of PPP and P3 software.
We thank CERN for the kind support and particularly S. Bertolucci, J. Serrano, P. Alvarez Sanchez and E. Gschwendtner. We finally thank G. Manusardi and F. Bastianini for valuable support with OTDR measurements.

\end{document}